\documentstyle[aaspp4]{article}

\begin{document}

\def\sec{$^{\prime\prime}$}
\def\min{$^{\prime}$}

\title{A revised and extended catalog of Magellanic System clusters, 
associations and emission nebulae. II. the LMC}

\author{Eduardo L. D. Bica\altaffilmark{1,2,3},
Henrique R. Schmitt\altaffilmark{1,2,3}, Carlos, M. Dutra\altaffilmark{1,2,3},
Humberto L. Oliveira\altaffilmark{1,2,3}}

\altaffiltext{1}{Depto. Astronomia, IF-UFRGS, CP\,15051, CEP\,91501-970,
     Porto Alegre, RS, Brazil}

\altaffiltext{2}{CNPq Fellow}

\altaffiltext{3}{emails:bica@if.ufrgs.br, schmitt@if.ufrgs.br, 
dutra@if.ufrgs.br, humberto@if.ufrgs.br}

\begin {abstract}

A survey of extended objects in the Large Magellanic Cloud was carried
out on the ESO/SERC R and J Sky Survey Atlases, checking entries in
previous catalogs and searching for new objects.  The census provided
6659 objects including  star clusters, emission-free associations and
objects related to emission nebulae. Each of these  classes contains 3
subclasses with intermediate properties, which are used to infer total
populations.  The survey includes cross-identifications among catalogs
and we present 3246 new objects.  We  provide  accurate positions,
classification, homogeneous measurements of sizes and position angles,
as well as information on cluster pairs and hierarchical relation for
superimposed objects. This unification and enlargement of catalogs is
important for future searches of fainter and smaller new objects. We
discuss the angular and size distributions of the objects of the
different classes. The angular distributions show two off-centered
systems with different inclinations, suggesting  that the LMC disk is
warped.  The present catalog together with  its previous counterpart
for the SMC and the inter-Cloud region provide a total population of
7847 extended objects in the Magellanic  System.  The angular
distribution of the ensemble reveals important clues on the interaction
between the LMC and SMC.

\end{abstract}


\section{Introduction}

The systematic study  of the Large Magellanic Cloud (LMC) properties is
nowadays possible partly owing to cataloging efforts carried out
throughout decades.  Hodge(1975) reviewed the history of  star cluster
catalogs in the LMC, providing a census  of 1614 objects at that time,
considering overlapping entries.  Hodge's (1960) catalog of red
clusters, Shapley \& Lindsay's (1963) overall catalog and Lyng\aa\ \&
Westerlund's (1963) catalog of clusters in the outer parts of the LMC
play a fundamental role in all aspects of research concerning these
objects.  Hodge \& Sexton (1966) provided an additional wide angle
general catalog for fainter clusters.

Hodge's (1980) deep plate survey sampled different LMC regions. He
predicted a total cluster population of $\approx$5100 objects, by
considering completeness effects.  Hodge (1988) studied additional
fields finding new clusters and, by taking into account a population of
2053 cataloged clusters and completeness effects, reevaluated the total
population down to $\approx$4200  clusters.

A technique similar to that used by Hodge (1988) was employed by
Kontizas, Metaxa \& Kontizas  (1988) in the LMC bar detecting new faint
clusters and studying their internal dynamical properties.  Several
other studies have contributed with new objects, and provided accurate
positions  for previously cataloged ones (Lauberts 1982; Olszewski et
al. 1988; Kontizas et al. 1990).

Stellar associations in the LMC have been cataloged and described by
Lucke \& Hodge (1970), which  has been a key reference for subsequent
studies.  In a recent  study of such objects Kontizas et al. (1994)
found new associations  in the LMC main body. Also, automated
techniques have detected  clusters in the north-east region of the LMC
(Bhatia \& MacGillivray 1989), and  associations in the inter-Cloud
region (Battinelli \& Demers 1992).

The  fundamental catalogs of emission nebulae in the LMC are Henize's
(1956), based on objective prism and direct-exposure plates, and that
by Davies, Elliot \& Meaburn (1976), based on deeper H$\alpha$ direct
plates.  Henize (1956) included planetary nebulae and stellar objects
which are beyond the scope of the present study. Otherwise, both
catalogs are dominated by HII regions. Some objects turned out to be
supernova remnants (SNR), which amounted to 32 in the LMC according to
Mathewson et al. (1985).  Two new supernova remnants have been recently
detected by Smith et al.  (1994).

This paper presents a unified deep catalog of star clusters, stellar
associations and emission nebulae in the LMC. It is a continuation of
the work started by Bica \& Schmitt (1995 -- Paper I), which was
devoted to the SMC and the inter-Cloud region (the stellar and H\,I
bridge linking the Clouds --- Irwin, Demers \& Kunkel 1990, Grondin,
Demers \& Kunkel 1992).  Homogeneous catalogs form a database useful
for systematic studies of  the structure and dynamics of the Clouds
(see Westerlund 1990 for a review), and for dating their objects in
view of recovering their star formation history (see e.g. Hodge 1973).

The paper is organized in the following way.  In Sect. 2  we present
the procedures employed in the construction of the catalog and show the
results, giving  examples of finding charts that can be built with the
present catalog.  In Sect.3 we discuss the angular distribution of the
different object types in the LMC, their relation to the HI
distribution, and also study their size distribution. In Sect.4 we use
the present LMC catalog together with that of the SMC and Bridge
regions (Paper I) to infer possible effects of the LMC and SMC
interaction.  The conclusions of this work are given in Sect.5.

\section {LMC Catalog}

In the 70's  the cataloged objects  were mostly contained in the LMC
Atlas (Hodge \& Wright 1967). In  Table 1 we  chronologically gathered
references before and after the LMC Atlas, indicate their corresponding
acronyms, number of entries and the object types which they basically
include (C, A or N for star cluster,  association and emission nebula,
respectively).  Discoveries of new objects have been numerous, but also
new designations for previous objects have been created and
cross-checking them was necessary.

The present revision also includes the LMC objects in the NGC and IC
catalogs, and the objects in the  ESO catalog (Lauberts 1982).  Some
catalogs of bright stars and/or of emission line stars  include some
compact star clusters (Heydari-Malayeri et al. 1988, 1989;
Heydari-Malayeri \& Hutsem\'ekers 1991; Heydari-Malayeri et al. 1993),
and compact HII regions (e.g. Heydari-Malayeri \& Testor 1983). Their
acronyms were used in the present study in order not to create new
designations:  HD or HDE, LMC-S -- Henize (1956); Sk -- Sanduleak
(1969).

  The objects were examined and measured on the ESO R and J Sky Survey
Schmidt films at the Instituto de F\'{\i}sica, UFRGS.  The measurements
for calculating accurate coordinates relative to reference stars, as
well as object sizes and position angles, were carried out with a 7x
magnifying lens equipped with a graduated glass providing  a length
resolution of 0.05 mm (which converts to less than 4 arcsec  on the
ESO  Schmidt films) and a precision in position angles of
$\pm$5$^{\circ}$.

  The procedures employed to built the LMC catalog follow those
  explained in detail in Paper I, so we briefly recall them in the
following: (i) digitalization of the original coordinates and
precession to the  epoch 1950 when necessary; (ii) identification of
the objects on the Schmidt films using all the available information in
terms of coordinates, descriptions, identification plates and finding
charts in the original papers and in the LMC Atlas (Hodge \& Wright
1967); (iii) equivalent objects cross-identified in different catalogs
were  merged into a single catalog line; (iv) transparent overlay
charts were generated via computer to  the same scale as the ESO
Schmidt films; (v) accurate coordinates, whenever available,  replaced
original ones, and for the rest, accurate positions were measured with
respect to reference stars; (vi) major and minor diameters, as well as
position angle of the major axis were measured.

 We illustrate in Table 2 the first five lines of the catalog; the
 whole catalog is available through the electronic version by link to
the University of Chicago Press archive. The Table is organized in the
following way. Field (1) gives the Sky Survey plate and quadrant where
the object is best seen.  Field (2) gives the object
cross-identification in the different catalogs; we have considered as
equivalent objects HII regions and their embedded stellar associations
of comparable extent; some objects were separated in two or more parts
and consequently the  designation is complemented with `n' for northern
part, `se' for southeastern part, etc.  We present 3246 new objects,
$\approx$ 49\% of the unified catalog.  In the SMC (Paper I) new
objects amounted to 24\% only, showing that previous surveys had
explored more efficiently the SMC, possibly because of its smaller
angular size.  Field (3) and (4) provide respectively right ascension
and declination precessed to J2000.0.  Field (5) indicates the object
type following  Paper I: the sequence C, CA, AC to A is essentially one
of density, where objects with the higher density of stars are
classified as C, and objects with the lower density as A.  Objects NA
and NC are stellar systems clearly related to emission, while N, in
this particular Table,  refers to known supernova remnants.  AN and CN
are stellar systems with traces of emission. Fields (6) and (7) give
the size of the object's major and minor axes, respectively. Field (8)
gives the position angle of the major axis (0$^{\circ}=$N,
90$^{\circ}=$E). Finally, Field (9) gives Remarks:  `mT' and `mP'
indicate member of triple or pair, respectively. A hierarchical
indication is given for objects embedded in or superimposed on larger
ones, where `in' suggests a possible physical connection while 'sup'
suggests a projection. Also, ``\&'' indicates additional
designation(s), which complement the catalog field  (2).

We show in Figures 1a and 1b two finding charts which can be built with
this catalog and IRAF task SKYMAP.  Figure 1a shows a dense region,
centered in the LMC bar, while Figure 1b shows a less crowded region,
located to the West of the bar.  These Figures identify objects of
different basic types (A, C and N) by different symbols, label them and
also include field stars for reference. The stars were obtained from
the Guide Star Catalog (Jenkner et al. 1990 and references therein).
Notice that some star clusters are also entries in the Guide Star
Catalog.

\subsection{Census}

We show in Table 3 the census of the subclasses of objects in the
present work. The most numerous objects are clusters and associations.
Including intermediate classifications, there are 2577 clusters
(C+CA+CN), 2883 emission--free associations (A+AC+AN) and 1199 emission
line related objects (N+NC+NA).  We basically confirm Hodge's (1988)
prediction of $\approx$4200 star clusters in the LMC, by adding to the
C+CA+CN the subclasses AC, consisting of somewhat looser objects, and
NC, which are newly formed star clusters, totaling 4089 objects.

Notice that the definition of association in the present study (and
also in Paper I) is related to the density of stellar systems, so they
are not necessarily young as the traditional OB associations,  but OB
associations (Lucke and Hodge 1970) are included  in this class as A,
NA or AN. The densities of stars in the lowest density objects  as
counted on Sky Survey plates are typically 5-10  stars per square
minute, a factor $\approx$2 larger than in surrounding regions.  They
often have irregular or elongated shapes.  Detection efficiencies
certainly  depend  on background star densities, and should become
critical in the bar region. In crowded regions, younger  objects
(higher surface brightness) are expected  to be  more easily detected
than older ones.

 As to the emission related objects, apart from the 35 known Supernova
Remnants (notice that LMC-DEM316 was separated into two components),
the remaining 1164 objects are mostly HII regions, with a wide range in
dimensions (Section 3.2)

\section{LMC}

\subsection{Angular Distribution}

Figures 2a, b, c and d show the angular distribution of: a) all
extended objects; b) clusters; c) associations; d) emission line
objects.  Superimposed on Figures 2b, c and d is the HI surface density
distribution in the LMC (Mathewson \& Ford 1984).  Figure 2a clearly
shows two stellar systems, a nearly circular dense internal disk with
dimensions $\approx6^{\circ}\times7^{\circ}$, and a less dense outer
disk with $\approx10^{\circ}\times14^{\circ}$.  The centers of the two
systems are separated by approximately 1$^{\circ}$.  Assuming that the
intrinsic distributions are circular, the axial ratios imply
inclinations of $\approx$45$^{\circ}$ and $\approx$30$^{\circ}$ for the
outer and inner systems, respectively. Such tilts would imply that the
LMC disk is warped. The line of nodes of the outer system is
approximately aligned with the N-S direction, while that of the inner
system is aligned along PA$\approx$130$^{\circ}$. The existence of two
disk systems (Figure 2a), can explain why estimates of geometrical and
kinematical parameters depend on the type of object used to determine
them (Westerlund 1990 and references therein).

In the southern outer region of the LMC there occurs  a structure in
the object distribution (Figs. 2, see also Fig. 4a), already described
by Lyng\aa\ \& Westerlund (1963), and  possibly related to a spiral arm
(de Vaucouleurs 1955). The increased number of objects cataloged in the
present study shows this structure in more detail. In order to test its
significance we counted objects in 1$^{\circ}\times$1$^{\circ}$ boxes
along the disk edge and slightly more internal  control zones. The box
centers, position angles, angular distances from the LMC bar center and
counts  are given  in Table 4. Considering $\sqrt{N}$ fluctuations for
the count $N$, it can be concluded that the southern structure is
statistically significant from position angles $\approx$140$^{\circ}$
to  $\approx$ 220$^{\circ}$.  Counts of field stars might reveal
whether this is an arm, a partial ring or part of a complete ring which
could be blurred by contamination in the eastern and northern regions,
and hidden by the off-centered high density inner disk to the west.
Irwin (1991)  presented LMC isopleths for field stars from
machine-based measurements of UKST plates.   He used horizontal
branch/giant clump stars, thus eliminating younger disk components. The
southern structure is clear, and an equally dense region is seen all
around the LMC, which could suggest the presence of a ring, but
unfortunately  in his Figure 3 adjacent control zones are available
only for the southern part.

The angular distribution of star clusters (Figure 2b) clearly shows the
LMC bar in the central region, which is not the case in the angular
distribution of associations (Figure 2c). This is due to the fact that
the stellar population of the bar has an important intermediate age
component (Hardy et al. 1984; Bica, Clari\'a \& Dottori 1992), and that
star formation in the last 3$\times 10^7$ years has been important in
the NE of the bar (Dottori et al. 1996), consequently enhancing the
number of OB associations in that region. In Figure 2b, the HI
distribution shares a similar center with the inner disk system and the
10$\times10^{19} atoms~ cm^{-2}$ contour does not encompass the outer
cluster distribution in the north east area. Such asymmetries between
the outer stellar disk and the HI distribution, again point to the
perturbed nature of the LMC.

In Figure 2c there is a strong concentration of associations in a
region depleted of HI at RA$\approx$ 5h 32min and
$\delta\approx-66^{\circ}$ 49$^{\prime}$. This region corresponds to
Shapley-III, where HI has probably been consumed and dispersed by winds
arising from intense star formation at approximately 10$^7$ years.

The ongoing star formation is  seen in the distribution of emission
line objects (Figure 2d), which is dominated by HII regions (some
Supernova Remnants are also present). The necessary HI surface density
for star formation is between 40 and 100$\times 10^{19} atoms~
cm^{-2}$, similar to the value found for the SMC in Paper I. The
densest HI distribution occurs to the south of the 30 Doradus complex.

Since our classification from clusters to associations is one of
density, some very loose objects are not necessarily young. The
distribution of the loose objects (Figure 2c) contains some quite far
from the center of the LMC. They could be older than typical
associations, corresponding to systems in the process of evaporation.
Indeed, Wielen (1971) found evidence that, in the Galaxy, the
dissolution of open clusters is an important phenomenon. One of the
possibilities to explain the gap of clusters in the range 4 to 9 Gyr in
the LMC (Da Costa 1991), is their dissolution (Geisler et al. 1997).
This effect is expected in a wide range of ages.  For example, Wielen
(1988) estimated that a typical dissolution time is $\approx$2 Gyr for
a cluster in the outer LMC with an initial mass of 500 M$_{\odot}$ and
a core radius of 1 pc.  It would be important to check by means of
color magnitude diagrams and luminosity functions, the age and
dynamical state of these outlying loose stellar systems.  Some loose
stellar systems  occur in the vicinity or are pairs with dense star
clusters (Table 2).  N--body simulations of star clusters  with
differing masses can produce as outcome loose dissolving systems from
the less massive component, depending on the collision conditions
(Oliveira, Dottori \& Bica 1998). The generalized present definition of
loose systems may emcompass objects originated by this mechanism. Loose
stellar systems are key objects to understand timescales and mechanims
by which star clusters and associations feed the field population.

\subsection{Size Distribution}

We show in Figs 3a, b and c the distributions of sizes (mean of major
and minor axes in Table 2)  respectively for the star clusters
(C+CN+CA), the associations (A+AN+AC) and the emission nebulae
(NA+NC+N). The largest star  clusters have log D(\min ) $\sim$0.7
(D=5\min ) and the number increases strongly towards smaller sizes down
to Log D(\min )$\sim$0.0 (D=1\min ). Subsequently there occurs a
plateau and finally a drop for Log D(\min )$<-$0.4 (D$\sim$24\sec),
where the number of objects in the bins is strongly affected by the
detection limit.

The size distribution of the associations (Figure 3b) shows a behavior
similar to that of the star clusters, but having some objects
considerably larger, with Log D(\min )$>$ 1 (D$>$10\min ). In the case
of the SMC and Bridge (Paper I), there is a sharp drop in the
distribution of sizes of associations at Log D(\min )$\approx0.7$,
which is due to the objects located in the Bridge. There is not such a
drop in the LMC histogram.  Notice the large population of small
associations (Log D(\min )$\sim-$0.2).  Many of these loose objects are
present in previous LMC cluster catalogs, as was the case also for the
SMC (Paper I).

The distribution of objects containing emission (Figure 3c) presents a
flatter distribution, which attains very large values, some of which
larger than 1$^{\circ}$ ($\approx$ 1 kpc). We point out that the latter
structures correspond to supergiant shells (Meaburn 1980) and Shapley
constellations (van den Bergh 1981), as well as large structures such
as LMC-N135 (Henize  1956).  Among the smallest objects there occur
compact HII regions with D$\le$1\min (e.g. Heydari-Malayeri \& Testor
1983).

\section{Magellanic system angular distribution}

The LMC catalog in conjunction with that of the SMC and Bridge regions
(Paper I), can be used to study the overall angular  distribution  of
extended objects in the Magellanic System, and their relation to the HI
distribution.

The total number of cataloged extended objects in the Magellanic System
amounts to 7847 entries: 3131 classified as clusters (C+CA+CN), 3226
associations essentially free of emission (A+AC+AN), and 1490 objects
related to emission (N+NA+NC). Figure 4a presents the angular
distribution of all objects, where we can clearly see the interacting
nature of the LMC and SMC, shown by the bridge of objects between the
two galaxies, as well as by the distortions occurring in the main
bodies and beyond.

We show in Figure 4b the angular distribution of star clusters, with
the Matthews \& Ford (1984) HI contours. Notice that clusters do not
trace the HI bridge, extending only as far as the SMC Wing.  As pointed
out in Section 3.1, the outer HI contour (10$\times10^{19} atoms~
cm^{-2}$) does not encompass the NE part of the LMC cluster
distribution, but encompasses essentially all the SMC clusters.  The
low HI contours are not symmetrically distributed with respect to the
cluster centroids, again denoting a perturbation in the HI and stellar
disks.

Figure 4c shows the distribution of associations, which follow the HI
distributions better than the star clusters, indicating that the nature
of the loose objects is mostly young. In particular, they populate the
HI bridge region. These objects are also abundant in the SMC wing and
in the main body of both galaxies. Finally, Figure 4d presents the
distribution of objects related to emission, which are mostly contained
inside higher HI contours (Section 3.1 and Paper I).  In the case of
the LMC we can see some structures, in particular the 30 Doradus
complex and concentrations related to Shapley constellations.  As
pointed out in Section 3.1, the LMC distribution of young objects
appears to be nearly face-on, while the distribution in the SMC is more
concentrated along an axis, which might be interpreted as a nearly
edge-on disk.  The 200$\times10^{19} atoms~ cm^{-2}$ HI contour in the
SMC is as well very elongated, also resembling an edge-on disk, with a
bump in the direction of the Wing, which appears to be present also in
the SMC nebulae distribution.

The above Figures show features possibly arising from the LMC and the
SMC interaction, whose basic characteristics  can be inferred from
available N-body simulations of a disk galaxy and a less massive
($\approx$ a factor 5) compact companion. Hernquist \& Weil (1993)
computed an axial collision, while   Quinn, Hernquist \& Fullagar
(1993) computed a collision  for a circular orbit inclined
30$^{\circ}$ relative to the primary galaxy disk plane  producing,
respectively, a  prominent outer ring, and a warping in the disk
component.  Both simulations produce disk thickening, and create  high
and low surface brightness components. Color-magnitude diagrams showed
evidence of depth effects in NE region of the LMC (Bica et al. 1998),
and  the present LMC results suggest a warping between a high and low
surface brightness disk components, together with a significant outer
structure in the southern part (Sect. 3.1). This simple comparison
suggests that an inclined  nearly axial  collision might reproduce the
basic structures seen in the LMC disk. A detailed  model should include
gas rich components for both Clouds, and the dominant Milky Way
potential. It should also describe the gas and young stellar bridge
between the Clouds (Irwin, Demers \& Kunkel 1990, Grondin, Demers \&
Kunkel 1992), and the distortions in the SMC (Paper I).

\section{Concluding Remarks}

We have unified the catalogs of star clusters, associations and
emission nebulae in the LMC. In the case of this revision we have
detected 3246 new objects, totaling 6659 entries in the general
catalog. We provide cross identifications among previous catalogs,
homogeneous measurements of positions, sizes and classification, which
is important for future studies of their properties and for the
identification of fainter and/or smaller objects.  The angular
distribution of object types is studied together with the HI
distribution.  Asymmetries are present in the LMC disk, which consists
of two off-centered systems with different inclinations, suggesting a
warping.  The previously identified  structure in the LMC southern edge
appears to be statiscally significant.  The present LMC catalog,
together with that of the SMC and Bridge regions (Paper I), provide a
total of 7847 objects, homogeneously cataloged.  The analysis of the
angular distribution of this ensemble provides interesting hints about
the interaction between the Clouds.

\acknowledgements We thank an anonymous referee for interesting
remarks. We thank   Miriani Pastoriza and Thaisa Storchi-Bergmann for
the efforts to acquire the ESO R and J Atlases to the Instituto de
F\'{\i}sica, and Horacio Dottori for the magnifying lens that we used
in the measurements.

\clearpage

\begin{figure}[h]
\caption[]{Examples of finding charts (J2000.0)
that can be generated with the present
catalog. (a) and (b) show high and low density regions, respectively. 
Plus signs identify clusters, crosses associations and diamonds nebulae.
The figures also include reference stars from the Guide Star Catalog,
identified by ellipse. The fainter stars shown are of magnitude 13 and 14.}
\end{figure}

\begin{figure}[h]
\caption[]{LMC angular distribution of (a) all objects; (b) star
clusters; (c) emission-free associations; (d) objects with emission. HI
surface density contours from Mathewson \& Ford (1984) for 10, 40, 100
and 200$\times10^{19} atoms~cm^{-2}$ are shown, increasing for outside
to inside.}
\end{figure}

\begin{figure}[h]
\caption[]{LMC size distributions of: (a) star clusters (C+CN+CA); 
(b) emission-free 
associations (A+AC+AN); (c) objects with emission (N+NA+NC).}
\end{figure}

\begin{figure}[h]
\caption[]{Magellanic System angular distributions of: (a) all objects;
(b) star clusters; (c) emission-free associations; (d) objects with
emission. HI surface density contours from Mathewson \& Ford (1984)
as in Figure 2. The declination tickmarks are separated by 10$^{\circ}$.}
\end{figure}

\end{document}